\documentclass[preprint,3p,times]{elsarticle}

\usepackage{amsfonts}
\usepackage{multirow}
\usepackage{mathrsfs}
\usepackage{graphicx}
\usepackage{amsmath}
\usepackage{amssymb}
\usepackage{bm}
\usepackage{bbm}
\usepackage{color}
\usepackage{hyperref}
\usepackage{slashed}
\usepackage{xparse}

\def\OMIT#1{}

\newcommand{\nn}{\nonumber}

\newcommand{\beq}{\begin{equation}}
\newcommand{\eeq}{\end{equation}}
\newcommand{\bqa}{\begin{eqnarray}}
\newcommand{\eqa}{\end{eqnarray}}

\newcommand{\bseq}{\begin{subequations}}
\newcommand{\eseq}{\end{subequations}}

\newcommand{\dd}{{\mathrm{d}}}
\newcommand{\ket}[1]{\left\lvert{#1}\right\rangle}

\newcommand{\braket}[2]{\left\langle{#1}\middle\vert{#2}\right\rangle}

\newcommand{\abs}[1]{\left\lvert{#1}\right\rvert}

\begin{document}



\title{Exclusive radiative production of fully-charmed tetraquarks at $B$ Factory}


\author[1,2]{Feng Feng}
\ead{F.Feng@outlook.com}

\author[1,3]{Yingsheng Huang}\ead{huangys@ihep.ac.cn}

\author[1,3]{Yu Jia}\ead{jiay@ihep.ac.cn}

\author[4]{Wen-Long Sang}\ead{wlsang@swu.edu.cn}

\author[1,3]{Jia-Yue Zhang}\ead{zhangjiayue@ihep.ac.cn}

\address[1]{Institute of High Energy Physics, Chinese Academy of
  Sciences, Beijing 100049, China\vspace{0.2 cm}}

\address[2]{China University of Mining and Technology, Beijing 100083, China\vspace{0.2 cm}}

\address[3]{School of Physics, University of Chinese Academy of Sciences,
  Beijing 100049, China\vspace{0.2 cm}}

\address[4]{School of Physical Science and Technology, Southwest University, Chongqing 400700, P.R. China}

\date{\today}

\begin{abstract}

  We apply the newly developed NRQCD factorization formula to analyze the exclusive production of the
  fully-charmed tetraquark $T_{4c}$ plus a hard photon at $B$ factory. For simplicity, in this work
  we have concentrated on
  the associated production of the $0^{++}$ and $2^{++}$ tetraquarks with a photon.
  With some rough estimates about the size of the unknown nonperturbative NRQCD matrix elements,
  we predict the production rates of these types of exclusive channels at $\sqrt{s}=10.6$ GeV
  and also assess their observation prospect at {\tt Belle 2} experiment.

\end{abstract}

\maketitle

\section{Introduction}

Very recently the {\tt LHCb} collaboration has reported the evidence of a narrow structure near $6.9\,\rm GeV$ in the di-$J/\psi$
invariant mass spectrum, with a global significance of more than $5\sigma$~\cite{Aaij:2020fnh}. This structure is also accompanied by a broad structure ranging from $6.2$ to $6.8\,\rm GeV$. It is widely believed that this narrow structure, also known as $X(6900)$, is a strong candidate for fully-charmed tetraquark (often abbreviated by $T_{4c}$).
This unexpected discovery appears to open a novel, clean and unique window for the study of exotic hadrons.
In sharp contrast to a plethora of candidates of exotic hadrons discoverer in past two decades (exemplified by the so-called $XYZ$ states)~\cite{Guo:2017jvc,Liu:2019zoy,Ali:2017jda,Brambilla:2019esw}), whose nature is
often blurred by the contamination due to light constitute
quarks, the $X(6900)$ appears to admit a most natural tetraquark interpretation, {\i.e.}, its leading Fock state can be unambiguously
assigned with $|cc\bar{c}\bar{c}\rangle$.
The underlying reason is simply because the charm quark is too heavy to be excited from the QCD vacuum.

Theoretical investigations on the fully heavy tetraquarks date back to late 1970s~\cite{Iwasaki:1976cn,Chao:1980dv}.
During the past four decades, a plenty of studies on the properties of the fully heavy tetrqaurk states have been scattered in literature.
For example, the mass spectra and decay patterns of $T_{4c}$ have been studied in  quark potential model~\cite{Bai:2016int,Berezhnoy:2011xn,Becchi:2020uvq,Debastiani:2017msn,Lu:2020cns},
as well as QCD sum rules~\cite{Wang:2017jtz,Chen:2020xwe,Wang:2020ols}.
Recently, more rigourous approach such as lattice NRQCD has also been applied to investigate the lowest-lying spectrum of the
$bb\bar b\bar b$ sector, but found no indication of any states below $2\eta_b$ threshold in the $0^{++}$, $1^{+-}$ and $2^{++}$ channels~\cite{Hughes:2017xie}.

Since the newly discovered $X(6900)$ lies far above the di-J/$\psi$ threshold, the possibility of being a loosely bound
molecule made of two $J/\psi$ can be immediately ruled out.
It could be argued that $X(6900)$ might be the molecular state composed of two excited charmonia.
Nevertheless, it is difficult to conceive the underlying mechanism that
the rather weak Van der Waals force between two quarkonia can effectively bind two charmonia to form a molecule.
The most appealing interpretation appears to view the $X(6900)$ particle as a compact fully-charmed tetraquark state.
Moreover, in the popular phenomenological diquark model, one may portray the $T_{4c}$ as the color-singlet state formed by
the diquark-antidiquark clusters.
Following this standpoint, Chen {\it et al.} interpret the narrow structure a $P$-wave tetraquark,
while the broad one being a $0^{++}$ tetraquark state, within the QCD sum rules approach~\cite{Chen:2020xwe}.
Alternatively, the $X(6900)$ particle is also interpreted as the first radical excitation of $0^{++}$~\cite{Lu:2020cns,Wang:2020ols,Karliner:2020dta}.

In contrast with the intensive studies on the mass spectra and decay properties,
there has only been scattered explorations on the production mechanism of the fully heavy tetraquarks.
Various phenomenological tools have been employed, {\it e.g.}, quark-hadron duality~\cite{Berezhnoy:2011xn,Becchi:2020uvq,Berezhnoy:2011xy,Becchi:2020mjz},
color evaporation model~\cite{Carvalho:2015nqf,Maciula:2020wri}, {\it etc.}.
Unfortunately, the numerical predictions in various model calculation
are often sensitive to the {\it ad hoc} assumptions on input parameters.

In fact, the physical picture underlying the production of fully heavy tetraquark is quite analogous to that underlying
the production of ordinary heavy quarkonium. The key observation is that, to form a $T_{4c}$ state,
four charm quarks have to be first created altogether at small spatial distance ($\le 1/m_c\ll 1/\Lambda_{\rm QCD}$),
which should proceed via a hard process, then hadronize into a physical
color-singlet state through some soft nonperturbative process.
Therefore, owing to asymptotic freedom, it is conceivable that the production rates for $T_{4c}$ states
can be written in a factorized form, which involves the product of
perturbatively-calculable short-distance coefficients (SDCs) and nonperturabtive long-distance matrix elements (LDMEs).
Motivated along this line of reasoning, very recently Ref.~\cite{Feng:2020riv} proposed a novel NRQCD factorization theorem for
inclusive production of the fully-heavy tetraquarks. As a concrete application, the authors of \cite{Feng:2020riv}
compute the $g\to T_{4c}(0^{++}/2^{++})$ fragmentation functions and estimate the production rates
of the fully heavy tetraquarks at large $p_T$ at {\tt LHC}.  
A similar NRQCD-inspired study for the medium-$p_T$ production of $0^{++}/2^{++}$ $T_{4c}$ at {\tt LHC} 
has also recently been conducted~\cite{Ma:2020kwb}.

In this work, our plan is to further apply the novel NRQCD factorization formalism developed in \cite{Feng:2020riv} to a
simpler environment, {\it e.g.}, the exclusive production of $T_{4c}$ in $e^+e^-$ annihilation rather than in $pp$ collision.
For simplicity, we will concentrate on the $0^{++}(2^{++})$ tetraquarks, which can be interpreted as $S$-wave diquark-antidiquark cluster. To ensure the $C$-conservation,
we are considering the associated production process $e^+e^-\to T_{4c}+\gamma$.
As a major phenomenological impetus, this research is spurred by the very high
luminosity of the recently commissioning {\tt Belle 2} experiment. We hope our study can shed some light on the
observation potentials of the fully charmed tetraquarks at $B$ factory.

The rest of the paper is organized as follows.
In Section~\ref{2}, we spell out the NRQCD factorization formula for exclusive production
of $T_{4c}+\gamma$, for both unpolarized and polarized differential cross section.
In Section~\ref{3}, we employ the perturbative matching method to deduce various NRQCD SDCs for $e^+e^-\to T_{4c}+\gamma$.
In Section~\ref{4}, adopting some reasonable ansatz for NRQCD LDMEs, we present the phenomenological predictions for
the production rates and assess the observation prospects of these channels.
In Section~\ref{5}, we summarize our results.
We devote \ref{app} to some detailed illustration on how to construct a perturbative/nonperturbativve
$T_{4c}$ state in the diquark picture, which plays the vital role for deducing 
the correct matching coefficients.

\section{NRQCD factorization for  exclusive production of $T_{4c}+\gamma$ in $e^+e^-$ annihilation\label{2}}

\begin{figure}
  \centering
  \includegraphics[width=0.8\textwidth]{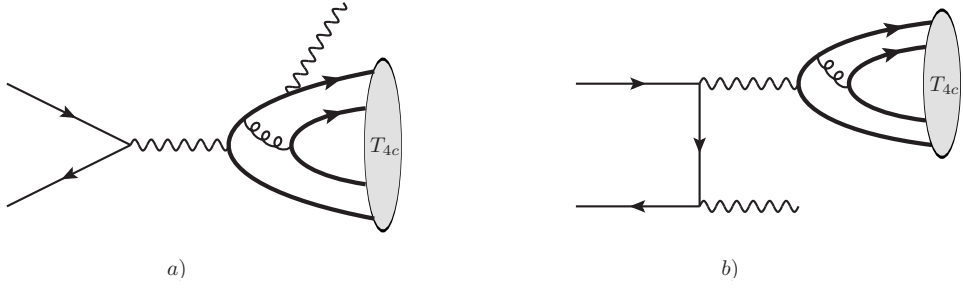}
  \caption{Characteristic Feynman diagrams for $e^+e^-\to T_{4c}+\gamma$. $a)$ designates the $s$-channel diagrams,
    while $b)$ signifies the $t$-channel diagrams.
    Since $C$-parity conservation forbids the $0^{++}(2^{++})$ tetraquark to
    be converted from a photon, so the category $b)$ does not contribute to our process. Moreover, due to parity invariance,
    the class $b)$ also does not contribute to radiative production of the $1^{+-}$ $S$-wave tetraqaurk.}
  \label{feyndiag}
\end{figure}

Our goal in this section is to introduce the NRQCD factorization formalism for $e^+e^-\to T_{4c}+\gamma$,
accurate at leading-order (LO) in velocity expansion.
For simplicity, in this work we will concentrate on the simplest fully-charmed $S$-wave tetraquarks carrying
the quantum number $0^{++}$ and $2^{++}$. Some typical Feynman diagrams are displayed in Fig.~\ref{feyndiag}.
Due to $C$-parity conservation, it is impossible for a photon to fragment into a $0^{++}(2^{++})$ tetraquark, so that the $t$-channel
process in Fig.~\ref{feyndiag}$b)$ does not contribute. Therefore we only need consider the $s$-channel diagrams
exemplified by Fig.~\ref{feyndiag}$a)$, which consists of roughly 40 diagrams in total.

Let us consider these exclusive production processes in the center-of-mass frame.
Assume the initial $e^\pm$ beams to be unpolarized, and let $\gamma$ and $T_{4c}$
carry the specific helicity $\lambda_1$ and $\lambda_2$, respectively.
For the production of $T_{4c}+\gamma$ via a virtual photon in $s$-channel $e^+e^-$ annihilation, one can
employ a standard shortcut, to express the differential polarized cross section in term of the
differential decay rate of a virtual photon:
\begin{align}
  \dfrac{\mathrm{d}\sigma\left[e^+e^-\rightarrow\gamma\left(\lambda_1\right)+T^J_{4c}\left(\lambda_2\right)\right]}{\mathrm{d}\cos\theta}=
   & \dfrac{2\rm{\pi}\alpha}{s^{3/2}}\sum_{S_z=\pm 1}\dfrac{\mathrm{d}\Gamma\left[\gamma^*\left(S_z\right)\rightarrow\gamma\left(\lambda_1\right)+T^J_{4c}
  \left(\lambda_2\right)\right]}{\mathrm{d}\cos\theta},
  \label{Shortcut:X:section:from:decay:rate}
\end{align}
where the polarizations of the $e^\pm$ in the initial state have been averaged.
$S_z$ is the spin projection of the virtual photon along the direction of the $e^-$ beams, $\theta$ is the polar
angle between the moving direction of $\gamma$ and the $e^-$ beam. Since the mass of $e^\pm$ has been neglected,
helicity conservation of QED demands the virtual photons to be transversely polarized, {\it i.e.}, $S_z=\pm 1$ in \eqref{Shortcut:X:section:from:decay:rate}.

In line with the standard helicity amplitude formalism,
The differential decay rate of a virtual photon into $\gamma(\lambda_1)+T^J_{4c}(\lambda_2)$
can be cast into
\begin{align}
  \dfrac{\mathrm{d}\Gamma\left[\gamma^*\left(S_z\right)\rightarrow\gamma\left(\lambda_1\right)+T^J_{4c}\left(\lambda_2\right)\right]}
  {\mathrm{d}\cos\theta}=\dfrac{\left|\mathbf{p}_f\right|}{16\rm{\pi} s}\dfrac{3}{4\rm{\pi}}\left|\mathcal{M}^J_{\lambda_1,\lambda_2}\right|^2\left|d^1_{S_z,\lambda}(\theta)\right|^2,
  \label{Decay:rate:from:hel:ampl}
\end{align}
where $\lambda\equiv\lambda_1-\lambda_2$, $|\mathbf{p}_f|=(s-M_{T^2_{4c}})/(2\sqrt{s})$
is the magnitude of the three-momentum of the $\gamma$($T_{4c}$) in the CM frame.
The angular distribution is solely governed by the Wigner rotation matrix $d^1_{S_z,\lambda}(\theta)$,
associated with the spin-1 irreducible representation of the $SU(2)$ group.
All the nontrivial strong interaction dynamics is encoded in the helicity amplitude $\mathcal{M}^J_{\lambda_1,\lambda_2}$,
which no longer bears any angular dependence.

Substituting \eqref{Decay:rate:from:hel:ampl} into \eqref{Shortcut:X:section:from:decay:rate}, we then arrive at the desired
formula for the differential polarized cross section:
\begin{align}
  \dfrac{\mathrm{d}\sigma\left[e^+e^-\rightarrow\gamma\left(\lambda_1\right)+T^J_{4c}\left(\lambda_2\right)\right]}{\mathrm{d}\cos\theta}=
   & \dfrac{3\alpha}{32\rm{\pi} s^2}\dfrac{\left|\mathbf{p}_f\right|}{\sqrt{s}}
  \left|\mathcal{M}^J_{\lambda_1,\lambda_2}\right|^2\sum_{S_z=\pm 1}\left|d^1_{S_z,\lambda}(\theta)\right|^2.
  \label{diff:pol:cross:section:hel:ampl}
\end{align}
Integrating \eqref{diff:pol:cross:section:hel:ampl} over the polar angle, and summing over all possible helicities in
the final states, we then obtain the integrated unpolarized cross section:
\begin{subequations}
  \label{integrated:unpol:cross:section:hel:ampl}
  \begin{align}
    \sigma\left[e^+e^-\rightarrow\gamma+T^{0}_{4c}\right]= & \dfrac{\alpha}{8\pi s^2}\dfrac
    {\left|\mathbf{p}_f\right|}{\sqrt{s}}\left(2\left|\mathcal{M}^0_{1,0}\right|^2\right),
    \\
    \sigma\left[e^+e^-\rightarrow\gamma+T^{2}_{4c}\right]= & \dfrac{\alpha}{8\pi s^2}\dfrac{\left|\mathbf{p}_f\right|}{\sqrt{s}}
    \left(2\left|\mathcal{M}^2_{1,0}\right|^2+2\left|\mathcal{M}^2_{1,1}\right|^2+2\left|\mathcal{M}^2_{1,2}\right|^2\right),
  \end{align}
\end{subequations}
where the factor of 2 inside the parenthesis takes into account of the contribution from the parity-flipped
helicity amplitudes.

Until now we have just recapitulated some well-known results, simply constrained by group theoretical consideration.
It is the time to invoke the NRQCD factorization to parameterize the production rates for these types of processes.
Similar to exclusive quarkonium production, the NRQCD factorization also holds true for exclusive
$T_{4c}+\gamma$ production at the helicity amplitude level:
\begin{align}
  \mathcal{M}^J_{\lambda_1,\lambda_2}= \frac{\mathcal{A}^{3[J]}_{\lambda_1,\lambda_2} }{ m^{4}} \sqrt{2 M_{T_{4c}}}
  \langle T_{4 c}^{J} \vert \mathcal{O}^{(J)}_{\mathbf{\bar 3}\otimes\mathbf{3}} \vert 0\rangle +
  \frac{\mathcal{A}^{6[J]}_{\lambda_1,\lambda_2} }{ m^{4}}
  \sqrt{2 M_{T_{4c}}} \langle T_{4 c}^{J}\vert \mathcal{O}^{(J)}_{\mathbf{6}\otimes\mathbf{\bar 6}} \vert 0\rangle+
  {\cal O}(v^2),
  \label{NRQCD:fac:helicity:ampl}
\end{align}
where $\mathcal{A}^{{\rm color}[J]}$ are the SDCs. Conventionally the $T_{4c}$ state in NRQCD matrix element admits
nonrelativistic normalization,   
so we insert a factor $\sqrt{2 M_{T_{4c}}}$ in front of NRQCD matrix element to compensate for the
relativistic normalization of the $T_{4c}$ state in the QCD amplitude in the left side of \eqref{NRQCD:fac:helicity:ampl}.

The color-singlet composite NRQCD operators $\mathcal{O}^{(J)}_{\rm{color}}$ ($J=0,2$) are defined as~\cite{Feng:2020riv}
\begin{subequations}
\begin{align}
& \mathcal{O}^{(0)}_{\mathbf{\bar 3}\otimes\mathbf{3}}=-\frac{1}{\sqrt{3}}[\psi_a^\dagger\sigma^i(i\sigma^2)\psi_b^*] [\chi_c^T (i\sigma^2)\sigma^i\chi_d]\;
\mathcal{C}^{ab;cd}_{\mathbf{\bar 3}\otimes\mathbf{3}},
\\
& \mathcal{O}^{(0)}_{\mathbf{6}\otimes\bar{\mathbf{6}}}=\frac{1}{\sqrt{6}}
[\psi_a^\dagger(i\sigma^2)\psi_b^*] [\chi_c^T(i\sigma^2)\chi_d]\;
\mathcal{C}^{ab;cd}_{\mathbf{6}\otimes\bar{\mathbf{6}}},
\\
& \mathcal{O}^{(2)kl}_{\mathbf{\bar 3}\otimes\mathbf{3}}=[\psi_a^\dagger\sigma^m(i\sigma^2)\psi_b^*] [\chi_c^T(i\sigma^2)\sigma^n\chi_d]\;
\Gamma^{kl;mn}
\;\mathcal{C}^{ab;cd}_{\mathbf{\bar 3}\otimes\mathbf{3}},
\end{align}
\label{NRQCD:composite:operators}
\end{subequations}
where $\psi^\dagger$ and $\chi$ are the Pauli spinor fields creating the $c$ and $\bar c$, respectively.
The Latin letters $a,\cdots, d=1,2,3$ are the color indices.
$\sigma^i$ ($i=1,2,3$) denote Pauli matrices.
The rank-$4$ cartesian tensor is defined by $\Gamma^{kl;mn}\equiv \frac{1}{2}(\delta^{k m} \delta^{l n}+\delta^{k n} \delta^{l m}-\frac{2}{3} \delta^{k l} \delta^{mn})$.
The subscript ``color'' indicates the color structure of the diquark fields and the antidiquark fields. For example, $\mathcal{O}_{\mathbf{
    \bar 3}\otimes\mathbf{3}}$ implies that the diquark field furnishes the
color anti-triplet $\mathbf{\bar 3}$ representation, and the anti-diquark fields furnish the color-triplet $\mathbf{3}$ representation.
Analogously, $\mathcal{O}_{\mathbf{6}\otimes\mathbf{\bar 6}}$ indicates that the diquark fields are in the color sextet $\mathbf{6}$ state,
while the anti-diquark in the color anti-sextet $\mathbf{\bar 6}$ state.
Concretely speaking, the desired color structure is projected by contracting with the following
color rank-4 tensors in the color space:
\begin{subequations}
  \ \begin{align}
     & \mathcal{C}^{ab;cd}_{\bar{\mathbf{3}} \otimes\mathbf{3}}\equiv \left(\sqrt{\frac{1}{2}}\right)^2 \epsilon^{abm}\epsilon^{cdn}\frac{\delta^{mn}}{\sqrt{3}}=\frac{1}{2\sqrt{3}}(\delta^{ac}\delta^{bd}-\delta^{ad}\delta^{bc})
    \\
     & \mathcal{C}^{ab;cd}_{\mathbf{6}\otimes\bar{\mathbf{6}}}
    \equiv \frac{1}{2\sqrt{6}}(\delta^{ac}\delta^{bd}+\delta^{ad}\delta^{bc}).
  \end{align}
  \label{color:tensor}
\end{subequations}
We remark that the NRQCD production operators in \eqref{NRQCD:composite:operators} form a complete set of basis at leading order in $v$,
which have a nonvanishing overlap with the $C$-even $S$-wave tetraquark states.

Plugging \eqref{NRQCD:fac:helicity:ampl} into \eqref{diff:pol:cross:section:hel:ampl} and \eqref{integrated:unpol:cross:section:hel:ampl},
we then obtain the differential and integrated cross sections for $e^+e^-\to T^J_{4c}+\gamma$. Concretely speaking, for the
total unpolarized cross sections, we have
\begin{align}
  \sigma(e^+e^-\to T^J_{4c}+\gamma)
   & = {\frac{F^{[J]}_{3,3} }{ m^8} (2 M_{T_{4c}}) }
  \left|\left\langle T_{4 c}^{(J)}\left|\mathcal{O}_{\mathbf{\bar 3}\otimes\mathbf{ 3}}^{(J)}\right|0 \right\rangle\right|^{2}+
  {\frac{F^{[J]}_{6,6} }{ m^{8}}}\; (2 M_{T_{4c}}) \left|\left\langle T_{4 c}^{(J)}\left|\mathcal{O}_{\mathbf{6}
   \otimes \bar{\mathbf{6}}}^{(J)}\right| 0\right\rangle \right|^{2}
  \nn                                                                                                                                                                                                                                                                                                                                       \\
   & +  {\frac{F^{[J]}_{3,6} }{ m^{8}}} \;(2 M_{T_{4c}}) 2{\rm Re}\left[\left\langle T_{4 c}^{(J)}
   \left|\mathcal{O}_{\mathbf{\bar 3}\otimes\mathbf{3}}^{(J)}\right| 0\right\rangle\left
   \langle  0\left|\mathcal{O}_{\mathbf{6} \otimes \bar{\mathbf{6}}}^{(J)\dagger} \right|T_{4 c}^{(J)}\right\rangle\right]+\cdots.
  \label{unpol:cross:section:NRQCD:fac}
\end{align}
where $F^{[J]}_{\rm color}$ represent the SDCs at the cross section level.
Substituting \eqref{NRQCD:fac:helicity:ampl} into \eqref{integrated:unpol:cross:section:hel:ampl},
comparing with \eqref{unpol:cross:section:NRQCD:fac},
we can directly read off the SDCs $F^{[J]}_{\rm color}$ in terms of the amplitude-level
SDCs $\mathcal{A}^{{\rm color}[J]}_{\lambda_1,\lambda_2}$ first introduced in \eqref{NRQCD:fac:helicity:ampl}:
\begin{subequations}
  \begin{align}
     & F^{[J]}_{3,3}=\frac{1}{2J+1}\dfrac{\alpha}{8\pi s^2}\dfrac{\left|\mathbf{p}_f\right|}{\sqrt{s}}\sum_{\lambda_1,\lambda_2}
    \left\vert {\mathcal A}^{3[J]}_{\lambda_1,\lambda_2}\right\vert^2,
    \\
     & F^{[J]}_{3,6}=\frac{1}{2J+1}\dfrac{\alpha}{8\pi s^2}\dfrac{\left|\mathbf{p}_f\right|}{\sqrt{s}}\sum_{\lambda_1,\lambda_2}
    \left({\cal A}^{3[J]}_{\lambda_1,\lambda_2}{\cal A}^{6[J]*}_{\lambda_1,\lambda_2}\right),
    \\
     & F^{[J]}_{6,6} =\frac{1}{2J+1}\dfrac{\alpha}{8\pi s^2}\dfrac{\left|\mathbf{p}_f\right|}{\sqrt{s}}
    \sum_{\lambda_1,\lambda_2} \left\vert {\cal A}^{6[J]}_{\lambda_1,\lambda_2}\right\vert^2.
  \end{align}
  \label{SDCs:FJ:in:term:of:AJ}
\end{subequations}

\section{SDCs via perturbative matching\label{3}}

The central task now is to determine the various NRQCD SDCs associated with $e^+e^-\rightarrow\gamma+T^J_{4c}$ at lowest order in
$v$ and $\alpha_s$. Since the SDCs are insensitive to long-distance physics,
we can employ the standard perturbative matching method, by replacing the physical tetraquark
state $T^J_{4c}$ with a free 4-quark state $\vert [cc][\bar{c}\bar{c}]\rangle$ in \eqref{NRQCD:fac:helicity:ampl}.
Computing both sides of \eqref{NRQCD:fac:helicity:ampl} in perturbative QCD and perturbative NRQCD, we are then
able to deduce the desired SDCs. To expedite the matching calculation, it is convenient to prepare the free 4-quark states
as the eigen-states of the angular momentum, which are constructed to bear the same quantum number $0^{++}$ and
$2^{++}$ as the physical tetraquark we are interested.

For simplicity, it is most convenient to decompose the free 4-quark states into a $S$-wave diquark-antiquark cluster. In \eqref{app} we elaborate on how we construct such ``fictitious" free 4-quark states bearing the definite $J^{PC}$.
The $0^{++}$ state can be constructed from two independent configurations of
$S$-wave diquark-antiquark cluster. The first configuration is that the $S$-wave diquark $\vert [cc]\rangle$ forms a color anti-triplet,
consequently a spin-triplet by Fermi statistics, similarly, the $S$-wave anti-diquark $\vert [\bar{c}\bar{c}]\rangle$
must form a color triplet and spin-triplet. When the diquark is orbiting with the antidiquark with $S$-wave, it is possible to form a
$0^{++}$ tetraquark state. The alternative configuration is that the $S$-wave diquark $\vert [cc]\rangle$ forms a color sextet,
consequently a spin-singlet by Fermi statistics, and the $S$-wave anti-diquark $\vert [\bar{c}\bar{c}]\rangle$ forms
a color anti-sextet and spin-singlet, so the $S$-wave diquark-antidiquark cluster is forced to bear the quantum number $0^{++}$.
By the rule of thumb for angular momentum addition,  one readily sees that, only the first configuration is feasible in order to form a $2^{++}$ tetraquark.

We start with computing the matrix element in the right hand side of \eqref{NRQCD:fac:helicity:ampl} in perturbative NRQCD.
From the NRQCD operators defined in \eqref{NRQCD:composite:operators}, together with
the ``fictitious'' tetraquark state $\mathcal{T}^{J,m_j}_{\rm{color}}$, which admit the
standard nonrelativistic normalization and are explicitly constructed in \eqref{T4c}, we obtain
\begin{subequations}
  \begin{align}
     & \left\langle \mathcal{T}^{0}_{\mathbf{\bar 3}\otimes\mathbf{3}}\left|\mathcal{O}_{\mathbf{\bar 3}\otimes\mathbf{3}}^{(0)}\right|0 \right\rangle= 4,
    \\
     & \left\langle \mathcal{T}^{0}_{\mathbf{6\otimes\mathbf{\bar {6}}}}\left|\mathcal{O}_{\mathbf{6}\otimes\mathbf{\bar{6}}}^{(0)}\right| 0\right\rangle = 2\sqrt{\frac{2}{3}},
    \\
   & \left\langle \mathcal{T}^{2,m_j}_{\mathbf{\bar 3}\otimes\mathbf{3}} \left|{\mathcal O}^{(2)ij}_{\mathbf{\bar 3}\otimes\mathbf{3}}  \varepsilon^{ij}(m_j) \right|  0 \right\rangle = 4.
  \end{align}
  \label{pLDME}
\end{subequations}

We proceed to calculate the left side of \eqref{NRQCD:fac:helicity:ampl}, the QCD amplitude $e^+e^-\rightarrow\gamma+T^J_{4c}$
at lowest order in $v$ and $\alpha_s$. We need replace the physical tetraquark $T^J_{4c}$ a fictitious one,
$\mathcal{T}^{J,m_j}_{\rm{color}}$ as constructed in \eqref{T4c}.
To enforce four outgoing charm quarks to have the desired orbital/spin/color
quantum number, we follow \cite{Feng:2020riv} to adopt a spin/color projector to expedite the calculation.
Concretely speaking, we first set all the momenta of charm quarks to be equal, then make the following replacement in
the QCD amplitude:
\begin{align}
  \bar u^a_i\bar u^b_j v^c_k v^d_l\to (\textsf{C}\Pi_\mu)^{ij}(\Pi_\nu \textsf{C} )^{lk}\mathcal{C}^{ab;cd}_{\rm color} J^{\mu\nu}_{0,1,2},
  \label{sub}
\end{align}
where the charge conjugate matrix $\textsf{C}\equiv i\gamma^2\gamma^0$, the color projection tensor
$\mathcal{C}^{ab;cd}_{\rm color}$ is defined in \eqref{color:tensor},
and the spinor projection tensors $J^{\mu\nu}_{0,1,2}$ are defined in \eqref{Jmu}.

After the aforementioned matching procedure, together with the projection of each individual helicity amplitude,
we then deduce the amplitude-level SDCs in \eqref{NRQCD:fac:helicity:ampl} to lowest order in $\alpha_s$:
\begin{subequations}
  \begin{align}
     & \mathcal{A}^{3[0]}_{1,0}=\mathcal{A}^{3[0]}_{-1,0}=
    -\dfrac{16{\rm{\pi}}^{5/2}\alpha\alpha_s\left(10-17r+9r^2\right)}{27\sqrt{3}(3-r)(2-r)},
    \\
     & \mathcal{A}^{6[0]}_{1,0}=\mathcal{A}^{6[0]}_{-1,0}=-\dfrac{16{\rm{\pi}}^{5/2}
      \alpha\alpha_s\left(10-9r+r^2\right)}{9\sqrt{3}(3-r)(2-r)},
    \\
     & \mathcal{A}^{3[2]}_{1,0}=\mathcal{A}^{3[2]}_{-1,0}=\dfrac{128{\rm{\pi}}^{5/2}\alpha\alpha_s}{27\sqrt{6}(3-r)},
    \\
     & \mathcal{A}^{3[2]}_{1,1}=\mathcal{A}^{3[2]}_{-1,-1}=\dfrac{512{\rm{\pi}}^{5/2}\alpha\alpha_s}{27\sqrt{2}(3-r)}
    \left(\frac{m }{ s^{1/2}}\right),
    \\
     & \mathcal{A}^{3[2]}_{1,2}=\mathcal{A}^{3[2]}_{-1,-2}=\dfrac{2048{\rm{\pi}}^{5/2}\alpha\alpha_s}{27(3-r)}
    \left(\frac{m }{ s^{1/2}}\right)^2.
  \end{align}
  \label{helicity:ampl:SDCs}
\end{subequations}
For notational abbreviation, we have introduced a dimensionless ratio $r\equiv \frac{16m^2}{ s}$.

From \eqref{helicity:ampl:SDCs}, \eqref{NRQCD:fac:helicity:ampl}, and \eqref{diff:pol:cross:section:hel:ampl},
one observes $\sigma[\gamma(\lambda_1)+T^J_{4c}(\lambda_2)]\propto s^{-1-|\lambda_1+\lambda_2|}$,
which is compatible with the celebrated helicity selection rule~\cite{Brodsky:1981kj}.

Plugging \eqref{helicity:ampl:SDCs} into \eqref{NRQCD:fac:helicity:ampl}, we deduce the analytic expression from each of
the helicity amplitude $\mathcal{M}^J_{\lambda_1,\lambda_2}$. We can use \eqref{diff:pol:cross:section:hel:ampl} to
predict the differential polarized cross section.
In order to predict the total unpolarized cross section as given in
\eqref{unpol:cross:section:NRQCD:fac}, we can directly use \eqref{SDCs:FJ:in:term:of:AJ}
to deduce the desired NRQCD SDCs $F^J_{\rm color}$:
\begin{subequations}
  \begin{align}
     & F^{[0]}_{3,3} =\frac{32\pi ^4 \alpha^3 \alpha_s^2(1-r)
      \left(10-17 r+9 r^2\right)^2 }{2187 s^2 (3-r)^2 (2-r)^2} ,
    \\
     & F^{[0]}_{3,6} =\frac{32\pi ^4 \alpha^3 \alpha_s^2(1-r)\left(10-9 r+r^2\right)\left(10-17 r+9 r^2\right)}
    {729 s^2\left(6-5 r+r^2\right)^2 },
    \\
     & F^{[0]}_{6,6}=\frac{32\pi ^4 \alpha^3 \alpha_s^2 (1-r)\left(10-9 r+r^2\right)^2 }
    {243 s^2(3-r)^2 (2-r)^2} ,
    \\
     & F^{[2]}_{3,3} =\frac{1024\pi ^4 \alpha^3 \alpha_s^2(1-r) \left(1+3 r+6 r^2\right) }
    {10935 s^2(3-r)^2} .
  \end{align}
  \label{F:J:explit:expressions}
\end{subequations}
Obviously, in the asymptotic limit $\sqrt{s}\gg m_c$, the production rate for $2^{++}$  plus photon is dominated by the
longitudinal polarization configuration, with $\sigma\propto s^{-2}$.

\section{Phenomenology on $T_{4c}+\gamma$ production at $B$ factory\label{4}}

We are now ready to assemble all the necessary ingredients in Sections~\ref{2} and \ref{3} together.
However, before exploring the phenomenology of $T_{4c}$ production at $B$ factory, we have to specify
the value of various input parameters, in particular, the value of the encountered nonperturbative NRQCD LDMEs
in \eqref{NRQCD:fac:helicity:ampl} and \eqref{unpol:cross:section:NRQCD:fac}.
In principle, the most reliable and systematic way to assess these LDMEs is through lattice NRQCD, which is,
unfortunately, not feasible at present.
In order to proceed, we resort to a simple phenomenological model, by treating the $T_{4c}$ as a diquark-antidiquark cluster.
We have argued in Sections~\ref{3} that the color decomposition of the $0^{++}$ tetraquark could be either $\mathbf{\bar 3}\otimes\mathbf{3}$ or $\mathbf{6}\otimes\mathbf{\bar 6}$ in the diquark-antidiquark basis,
whereas that of the $2^{++}$ $T_{4c}$ is solely of the form $\mathbf{\bar 3}\otimes\mathbf{3}$.
Furthermore, since the interquark short-range Coulomb force inside the color-sextet diquark is repulsive,
we decide to neglect the contribution of the $\mathbf{6}\otimes\mathbf{\bar 6}$ Fock component
in the $0^{++}$ tetraquark.
Nevertheless, we emphasize that {\it a priori} there is no fundamental reason to exclude the color-sextet contribution,
and our deliberate omission
of this term is entirely driven by simplicity consideration.

In \ref{app} we explicitly construct the $0^{++}$ and  $2^{++}$ states in the picture of diquark-antidiquark cluster.
We introduce two nonperturbative wave functions, {\it e.g.}, $R_D(r)$ signifies the radial Schr\"{o}dinger wave function
of the diquark/antidiquark, and $R_T(R)$ represents the radial wave function of the tetraquark by
treating the diquark and antidiquark as point particles.
From \eqref{NRQCD:composite:operators} and the tetraquark states constructed in \eqref{Tetraquark:state:construction},
we can explicitly express the NRQCD vacuum-to-$T^J_{4c}$ matrix element as the triple production of the
wave functions at the origin~\cite{Feng:2020riv}~\footnote{In a recent work about inclusive fully-charmed
tetraquark production at {\tt LHC}~\cite{Ma:2020kwb}, the authors chose the
nonperturbative parameter in their factorization formula
as the single wave function at the origin from solving the four-body Schr\"{o}dinger equation.
The connection between their nonperturbative factors and our approximated expressions for the NRQCD
matrix elements in \eqref{NRQCD:vacuum:T4c:matrix:element} appears not straightforward.}:
\bseq
\begin{align}
  \left\langle T_{4c}^{0} \left|\mathcal{O}_{\mathbf{\bar 3}\otimes\mathbf{3}}^{(0)}\right|0\right\rangle
  & \approx \frac{1}{2\pi^{3 / 2}} R_{\mathcal{D}}^{2}(0) R_{T}(0),
  \label{me0}
  \\
  \left\langle T_{4c}^{2,m_j} \left|{\mathcal O}^{(2) kl}_{\mathbf{\bar 3}\otimes\mathbf{3}} \varepsilon^{kl}(m_j) \right| 0\right\rangle & \approx
  \frac{1}{2\pi^{3/2}} R_{\mathcal{D}}^{2}(0) R_{T}(0) ,
\end{align}
\label{NRQCD:vacuum:T4c:matrix:element}
\eseq
where $m_j$ denotes the magnetic number of the $2^{++}$ tetraquark,
and $\varepsilon^{kl}(m_j)$ is the corresponding polarization tensor.

Squaring the vacuum-to-$T_{4c}$ matrix element and summing over the polarization,
we then obtain the intended NRQCD LDMEs appearing in the factorization formula for the production cross section
\eqref{unpol:cross:section:NRQCD:fac}:
\bseq
\begin{align}
  \left|{\left\langle {T_{4c}^{0}}\left|{\mathcal{O}_{\mathbf{\bar 3}\otimes\mathbf{3}}^{(0)}}\right|0\right\rangle}\right|^2 & =\frac{1}{4\pi^3}
  \left|{R_\mathcal{D}(0)}\right|^4\left|{R_T(0)}\right|^2,
  \\
  \sum_{m_j} \left|{\left\langle {T_{4c}^{2,m_j}}\left|{\mathcal{O}_{\mathbf{\bar 3}\otimes\mathbf{3}}^{(2)kl}}
  \right|0\right\rangle}\right|^2
  & =\frac{5}{4\pi^3}\left|{R_\mathcal{D}(0)}\right|^4\left|{R_T(0)}\right|^2.
\end{align}
\label{Squared:NRQCD:LDME:wave:fun}
\eseq
Up to the factor of 5, which reflects the difference between the number of polarizations of the
$0^{++}$ and $2^{++}$ tetraquarks, the two matrix elements in \eqref{Squared:NRQCD:LDME:wave:fun} are identical.
This can be viewed as a manifestation of approximate heavy diquark spin symmetry.

In phenomenological analysis, we choose $R_{\mathcal{D}}(0)=0.523\;\mathrm{GeV}^{3/2}$~\cite{Kiselev:2002iy},
and $R_T(0)=2.902\;\mathrm{GeV}^{3/2}$ resulting from Cornell-type potential model~\cite{Debastiani:2017msn}.
We set $m_c=1.5\,\rm GeV$, $\alpha_{\rm EM}\left(10.58\,\rm GeV\right)=1/130.9$~\cite{Bodwin:2007ga},
$\alpha_s(2m_c)=0.2355$~\cite{Chetyrkin:2000yt}.

\begin{figure}[!hbtp]
  \centering
  \includegraphics[width=0.7\textwidth]{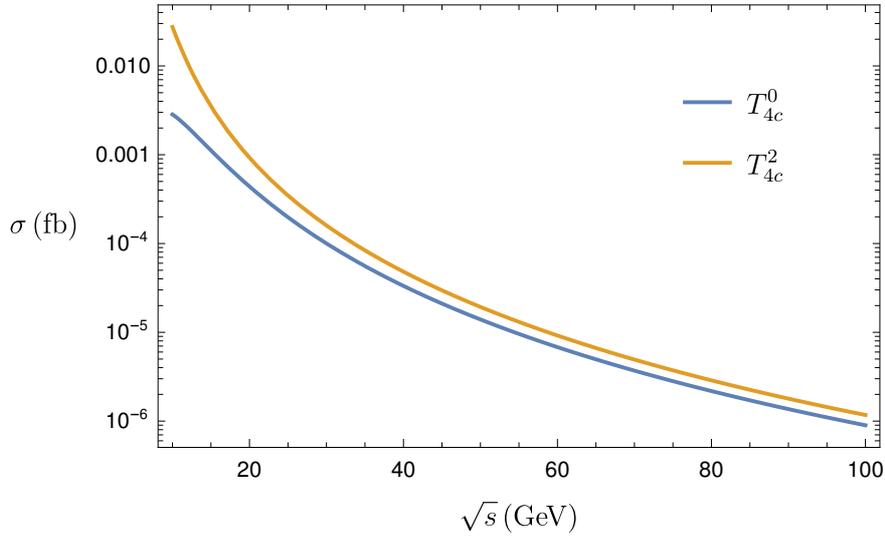}
  \caption{The total cross sections of $e^+e^-\rightarrow\gamma+T_{4c}^J\:(J=0,2)$
  as a function of the center-of-mass energy.}
  \label{fig:upcs}
\end{figure}

In Fig.~\ref{fig:upcs} we plot the production rates of $e^+e^-\rightarrow\gamma+T^J_{4c}$ ($J=0,2$) as a function
of the CM energy. One can clearly see that the total cross sections for these exclusive processes
decline quite fast with increasing $\sqrt{s}$,
in fact, $\sigma \propto 1/s^2$ as indicated in \eqref{F:J:explit:expressions}.

\begin{figure}[!hbtp]
  \centering
  \includegraphics[width=0.8\textwidth]{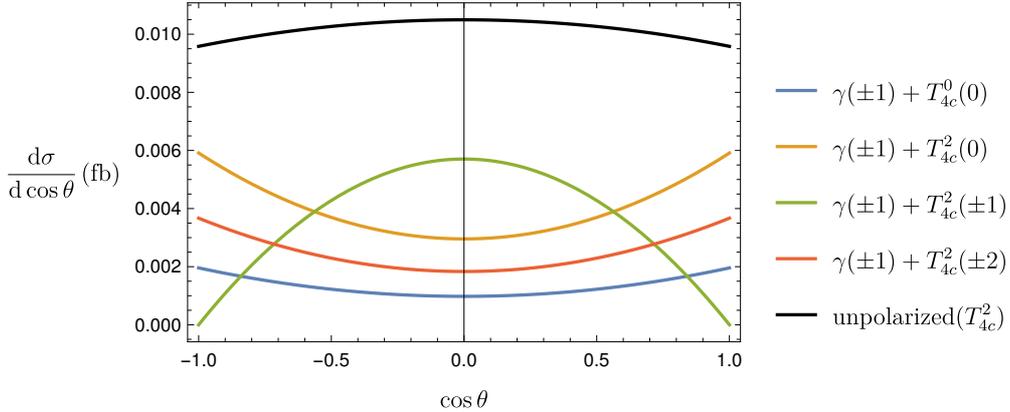}
  \caption{The angular distributions of $T_{4c}^J$ ($J=0,2$) at $\sqrt{s}=10.58$ GeV. Both polarized and
  unpolarized cases are juxtaposed. For each polarized cross section, we have also included the contribution from the
  helicity-flipped channel.}
  \label{fig:pcs}
\end{figure}

Let us specialize to the $B$ factory energy. Choosing $\sqrt{s}=10.58$ GeV, we find
\begin{subequations}
  \begin{align}
    \sigma\left[e^+e^-\rightarrow T_{4c}^0+\gamma \right]  & \approx  0.0026\;{\rm fb},
    \\
    \sigma\left[e^+e^-\rightarrow T_{4c}^2+\gamma  \right] & \approx  0.020\;{\rm fb}.
  \end{align}
\end{subequations}
Though being very much suppressed, the production rate for $2^{++}+\gamma$ is roughly one order of magnitude greater than that for $0^{++}+\gamma$,
compatible with what is found about $T_{4c}$ fragmentation production at {\tt LHC}~\cite{Feng:2020riv}.
It is interesting to compare with the exclusive radiative production of ordinary $C$-even quarkonium.
At LO in $\alpha_s$ and $v$, the production rate for $e^+e^-\rightarrow \eta_c + \gamma$ at $B$ factory is predicted to be $82\pm 21$ fb, and
that for $e^+e^-\rightarrow \chi_{cJ} + \gamma$ at $B$ factory are about $1.3\pm 0.2$ fb, $13.7\pm 3.4$ fb, and $5.3\pm 1.6$ fb for $J=0,1,2$, respectively~\cite{Chung:2008km}.
Several orders of magnitude suppression of the $T_{4c}$ production compared with ordinary charmonium is quite
conceivable, partly due to the severe penalty paid by producing two extra charm quarks,
and partly due to smaller probability for four charm quark to coincide in space.

The angular distributions of $T^J_{4c}$ from various helicity configurations,
together with the unpolarized differential cross section, are illustrated in Fig.~\ref{fig:pcs}.

The recently commissioning {\tt Belle 2} experiment is designed to have an
integrated luminosity about $50\;\mathrm{ab}^{-1}$\cite{Kou:2018nap}. Therefore,
in the full running period, we estimate that there are in total about $130$
$\gamma+T_{4c}^0$ events and $1020$ $\gamma+T_{4c}^2$ events at {\tt Belle 2} experiment.
If the $X(6900)$ particle is one of the $T_{4c}$ states we studied, it can be tagged through decay chains $X(6900)\to 2J/\psi(\to l^-l^+)$.
The small branching fraction of leptonic decay of $J/\psi$ together with detection efficiency
render the observation prospect of $X(6900)+\gamma$ events at {\tt Belle 2} extremely pessimistic.

Nevertheless, we should emphasize that, our estimate of the NRQCD matrix elements in
\eqref{NRQCD:vacuum:T4c:matrix:element} based on diquark model is very rough. Therefore,
our numerical results should be viewed as the exploratory in nature.
Moreover, the utter neglect of the color-sextet contribution may not be justified at all.
Under certain assumptions, it was shown that the color-sextet channel actually may
yield significant contribution for the $T^0_{4c}$ production rate at {\tt LHC}~\cite{Ma:2020kwb}.
In any rate, more reliable estimation of all the encountered NRQCD matrix elements are compulsory for an
accurate prediction for $T_{4c}+\gamma$ production.

\section{Summary\label{5}}

In this work, we have made a comprehensive study on the associated production of the
fully-charmed tetraquark $T_{4c}$ associated with a hard photon at $B$ factory.
We have concentrated on two simplest tetraquarks with quantum number $0^{++}$ and $2^{++}$,
which can be classified as the $S$-wave diquark-antidiquark cluster.
The calculation is based on the newly developed NRQCD factorization formalism by the authors,
and the short-distance coefficients have been calculated to lowest order in velocity and $\alpha_s$.
We employ the diquark model to make some very crude estimates about the magnitude about the hitherto
unknown nonperturbative NRQCD matrix elements. We find that
the production rates at $\sqrt{s}=10.6$ GeV are too small for these exclusive
channels to be observed at {\tt Belle 2} experiment.
Model-independent estimates on the NRQCD matrix elements are required to make more
reliable predictions for the exclusive radiative production of $T^J_{4c}$ at $B$ factory.
\\

{\noindent\bf Acknowledgment.}
The work of F.~F. is supported by the National Natural
Science Foundation of China under Grant No. 11875318,
No. 11505285, and by the Yue Qi Young Scholar Project
in CUMTB.
The work of Y.-S.~H., Y.~J. and J.-Y.~Z. is supported in part by the National Natural Science Foundation of China under Grants No.~11925506, 11875263, No.~11621131001 (CRC110 by DFG and NSFC).
The work of W.-L. S. is supported by the National Natural Science Foundation
of China under Grants No. 11975187 and the Natural Science Foundation of ChongQing under Grant No. cstc2019jcyj-msxm2667.

\appendix

\section{Constructing the tetraquark states and NRQCD matrix elements\label{app}}

In order to determine the SDCs from pertrubative matching, 
we need to construct a fictitious tetraquark made of free $cc\bar c \bar c$ quarks. 
We define the perturbative diquark state $\ket{[cc]}$ in certain angular momentum eigen-state as $\ket{\mathcal{D}_k^{m_j}(P)}$. 
We label the free diquark/tetraquark states by the calligraphy letters $\mathcal{D}$ and $\mathcal{T}$.
Later when we express the nonperturbative NRQCD matrix elements in terms of the wave functions at the origin, 
we will refer to the bound diquark/tetraquark states by the label $D$ and $T$.

Let us first consider the $S$-wave diquark in the spin-triplet/color anti-triplet channel. 
Since in this work we work only at the lowest order
in $v$, we are justified to neglect the relative motion between two $c$ quarks inside the diquark. 
Bearing the momentum $P$, the diquark state can be built as 
\begin{align}
\ket{\mathcal{D}_d^{s}(P)}=\frac{1}{\sqrt{2}} \sum_{\lambda_1\lambda_2}\braket{\frac{1}{2}\lambda_1\frac{1}{2}\lambda_2}{1s}\frac{\epsilon^{abd}}{\sqrt{2}}
\ket{c_a^{\lambda_1}\left({P\over 2}\right) c_b^{\lambda_2}\left({P\over 2}\right)},
\label{Di}
\end{align}
where $\braket{j_1 m_1 j_2 m_2}{J m_j}$ is Clebsch-Gordan coefficient, 
$\lambda_i$ ($i=1,2$) marks the magnetic numbers of the $c$ quarks.
$s$ and $d$ are the magnetic number and color index of the diquark.
The prefactor $1/\sqrt{2}$ is introduced to account for the two identical quarks. 
The anti-diquark state $\ket{\overline{\mathcal{D}}_m^{s}(P)}$  can be constructed in similar fashion.

Following similar procedure, we can combine the diquark and anti-diquark into a $S$-wave tetraquark state denoted by ${}^{2J+1}S_J$:
\begin{align}
   & \ket{\mathcal{T}^{J,m_j}(Q)}=\sum_{s_1s_2}\braket{1s_11s_2}{Jm_j}\frac{\delta^{de}}{\sqrt{3}}\ket{D^{s_1}_d(P)}\ket{\overline{D}^{s_2}_e(Q-P)}.
\end{align}
Substituting the explicit expression of $\ket{\mathcal{D}_d^{s}(P)}$ in \eqref{Di} into this equation, 
we can construct the $S$-wave tetraquark states in both $\mathbf{\bar 3}\otimes\mathbf{3}$ and $\mathbf{6}\otimes\mathbf{\bar 6}$ channels 
as follows:
\begin{subequations}
  \begin{align}
     & \ket{\mathcal{T}^{J,m_j}_{\mathbf{\bar 3}\otimes\mathbf{ 3}}(Q)}=\frac{1}{2}\sum_{\substack{s_1s_2                                                                                                                                                                                                                                                                                  
     \\
     \lambda_1\lambda_2\lambda_3\lambda_4}}\braket{\frac{1}{2}\lambda_1\frac{1}{2}\lambda_2}
     {1s_1}\braket{\frac{1}{2}\lambda_3\frac{1}{2}\lambda_4}{1s_2}\braket{1s_11s_2}{Jm_j}\mathcal{C}_{\mathbf{3} \otimes{\mathbf{3}}}^{ab;cd}\ket{c_a^{\lambda_1}(q_1)c_b^{\lambda_2}(P-q_1)\bar{c}_c^{\lambda_3}(q_2)\bar{c}_d^{\lambda_4}(Q-P-q_2)}\\
     & \ket{\mathcal{T}^{0,0}_{\mathbf{6}\otimes\mathbf{\bar 6}}(Q)}=\frac{1}{2}\sum_{\lambda_1\lambda_2\lambda_3\lambda_4}\braket{\frac{1}{2}\lambda_1\frac{1}{2}\lambda_2}{00}\braket{\frac{1}{2}\lambda_3\frac{1}{2}\lambda_4}{00}\mathcal{C}_{\mathbf{6} \otimes\bar{\mathbf{6}}}^{ab;cd}\ket{c_a^{\lambda_1}(q_1)c_b^{\lambda_2}(P-q_1)\bar{c}_c^{\lambda_3}(q_2)\bar{c}_d^{\lambda_4}(Q-P-q_2)},
  \end{align}
\label{T4c}
\end{subequations}
where $Q$ is the momentum of the tetraquark, $\mathcal{C}_{\rm{color}}^{ab;cd}$ is the color projection tensor given in 
\eqref{color:tensor}. Note that the $\mathbf{6}\otimes\mathbf{\bar 6}$ case only accommodate the $J=0$ state. 
Armed with these explicitly constructed tetraquark states, working in the rest frame, 
we can readily reproduce the perturbative NRQCD LDMEs in \eqref{pLDME}.

We should also use the same tetraquark state \eqref{T4c} in the perturbative QCD-side calculation.
As mentioned in Section~\ref{2}, in order to expedite the calculation, we can make use of the covariant
spin projector widely used in quarkonium community~\cite{Petrelli:1997ge}:
\begin{subequations}
\begin{align}
 v(P/2)\bar{u}(P/2) & \longrightarrow \Pi_0(P)            
  =\frac{1}{2 \sqrt{2}}\gamma^5\left(\slashed P+2m\right),                                                                                                                                                                                                                                                                                            
  \\
 v(P/2)\bar{u}(P/2) & \longrightarrow \Pi^\mu (P) =\frac{1}{2 \sqrt{2}}\slashed\gamma^\mu \left(\slashed P+2m\right),
\end{align}
\end{subequations}
where $P$ is the total momentum of $c\bar c$ system.

For the tetraquark production in the QCD side, there are four outgoing charm quarks, which correspond to the 
product of four Dirac spinors $\bar u\bar u vv$. In order to project the $cc$ into an desired $S$-wave diquark, 
we can use the charge conjugate matrix $\textsf{C}$ to convert one $\bar u$ spinor into a $v$ spinor by the identity
$\bar u=v^T \textsf{C}$, and similarly convert one $v$ spinor to $\bar u$ by the identity $v={\textsf C} \bar{u}^T$. 

Since the two spin-$1$ diquarks are in relative $S$ wave, they can combine into a tetraquark with $J=0, 1,2$. 
We borrow the spin-orbital projector from \cite{Braaten:2002fi} and convert it into the spin-spin projector $J^{\mu\nu}_{0,1,2}$ in \eqref{sub}:
\begin{subequations}
  \begin{align}
  J_{0}^{\mu \nu}           & =\frac{1}{\sqrt{3}} \eta^{\mu \nu}(P)                                                                                                                                                             \\
    J_{1}^{\mu \nu}(\epsilon) & =-\frac{i}{\sqrt{2 P^{2}}} \varepsilon^{\mu \nu \rho \sigma} \epsilon_{\rho} P_\sigma,                                                                                                                      \\
    J_{2}^{\mu \nu}(\epsilon) & =\epsilon_{\rho \sigma}\left\{\frac{1}{2}\left[\eta^{\mu \rho}(P) \eta^{\nu \sigma}(P)+\eta^{\mu \sigma}(P) \eta^{\nu \rho}(P)\right]-\frac{1}{3} \eta^{\mu \nu}(P) \eta^{\rho \sigma}(P)\right\},
  \end{align}\label{Jmu}
\end{subequations}
with 
\begin{align}
    \eta^{\mu \nu}(P)         & =-g^{\mu \nu}+\frac{P^{\mu} P^{\nu}}{P^{2}}                                                                                                                                                      \end{align}

In order to compute the the nonperturbative NRQCD LDMEs in phenomenological analysis, we construct the diquark and tetraquark states in potential model,
endowed with Schr\"{o}dinger wave functions $\psi_D(\bm{r})=R_D(r)Y_{00}(\hat{\bm{r}})$ and $\psi_D(\bm{r})=R_T(r)Y_{00}(\hat{\bm{r}})$ (since we do not consider the effect of $\mathbf{6}\otimes\mathbf{\bar 6}$ color structure, in the following we assume $\mathbf{\bar 3}\otimes\mathbf{ 3}$ to be the default setting), 
such that for the diquark state
\begin{align}
  \ket{D_k^{m_j}(P)}=\frac{1}{\sqrt{2}}\int \frac{q^2\dd q}{(2\pi)^3}\tilde R_D(q)\int\dd \Omega_q Y_{00}(\hat{\bm{q}})\sum_{\lambda_1\lambda_2}\braket{\frac{1}{2}\lambda_1\frac{1}{2}\lambda_2}{1m_j}
  \frac{\epsilon^{ijk}}{\sqrt{2}}\ket{c_i^{\lambda_1}(q)c_j^{\lambda_2}(P-q)},
  \label{DiR}
\end{align}
where $Y_{lm}(\hat{\bm{q}})$ is the spherical harmonics, and $\tilde R_D(q)$ is the diquark 
radial wave function in momentum space. The radial wave functions follow the normalization convention:
\begin{align}
  \int\frac{q^2\dd q}{(2\pi)^3}\abs{\tilde R(q)}^2=1.
\end{align}
Utilizing some identities about Clebsch-Gordan coefficients such as
\begin{align}
  \sum_{\lambda_1\lambda_2}\braket{s_1\lambda_1s_2\lambda_2}{Jm_j}\braket{s_1\lambda_1s_2\lambda_2}{J'm_j'}=\delta^{JJ'}\delta^{m_jm_j'},
\end{align}
we can readily verify that the diquark state indeed satisfies the standard nonrelativistic normalization:
\begin{align}
  \braket{D^{s'}_i(P_1)}{D^{s}_j(P_2)}=(2\pi)^3\delta^{(3)}(\bm{P_1}-\bm{P_2})\delta^{ss'}\delta^{ij}.
\end{align}

For tetraquark state, we also implement the wave function $\psi_\mathcal{T}$ between the diquark and antidiquark:
\begin{align}
   & \ket{T^{J,m_j}(Q)}=\int \frac{P^2\dd P}{(2\pi)^3}\tilde R_T(P)\int\dd \Omega_P Y_{00}(\hat{\bm{P}})\sum_{s_1s_2}\braket{1s_11s_2}{Jm_j}\ket{D^{s_1}(P)}\ket{D^{s_2}(Q-P)}
  \label{Tetraquark:state:construction}
\end{align}
which satisfies the non-relativistic normalization condition
\begin{align}
   & \braket{T^{J',m_j'}(Q_1)}{T^{J,m_j}(Q_2)}=(2\pi)^3\delta^{(3)}(\bm{Q_1}-\bm{Q_2})\delta^{JJ'}\delta^{m_jm_j'}.
\end{align}

We can employ this nonperturbative tetraquark state to deduce the value of the LDMEs in \eqref{unpol:cross:section:NRQCD:fac}. 
Take $\left\langle 0\left|\mathcal{O}_{\mathbf{\bar 3}\otimes\mathbf{3}}^{(0)}\right| T_{4 c}^{0}\right\rangle$ as an example. 
Contracting the $c$ quarks and anti-quarks in the tetraquark state with the NRQCD operators in $\mathcal{O}_{\mathbf{\bar 3}\otimes\mathbf{3}}^{(0)}$, we immediately arrive at
\begin{align}
  \left\langle 0\left|\mathcal{O}_{\mathbf{\bar 3}\otimes\mathbf{ 3}}^{(0)}\right| T_{4 c}^{0}\right\rangle=\int \frac{q_1^2q_2^2P^2\dd q_1\dd q_2\dd P}{2^6\pi^{\frac{15}{2}}}\tilde R_D(q_1)\tilde R_D(q_2)\tilde R_T(P)S^{\lambda_1\lambda_2\lambda_3\lambda_4}{\cal C}^{ab;cd}_{\mathbf{\bar 3}\otimes\mathbf{3}},
\end{align}
where $S^{\lambda_1\lambda_2\lambda_3\lambda_4}$ contains all the spin information in the form of Clebsch-Gordan coefficients.  
We now utilize the fact that $\tilde R_\mathcal{D/T}(q)$ is the Fourier transform of the coordinate space radial wave function $R_\mathcal{D/T}(r)$:
\begin{align}
  R_\mathcal{D/T}(0)=\int\frac{q^2\dd q}{(2\pi)^3}\tilde R_\mathcal{D/T}(q).
\end{align}
With some basic algebra, we can recover \eqref{me0}.


\end{document}